\documentclass[a4paper,11pt]{article}
\usepackage{pos}

\title{Numerical GR MHD simulations of the post-merger system with a composition-dependent equation of state}

\author*[a]{Agnieszka Janiuk}
\author[a]{Gerardo Urrutia}

\affiliation[a]{Center for Theoretical Physics, Polish Academy of Sciences\\
  Al. Lotnikow 32/46, Warsaw, Poland}

\emailAdd{agnes@cft.edu.pl}

\abstract{By means of HARM\_COOL\_EOS, which is our code for conservative relativistic magnetohydrodynamics, we developed a new scheme for the simulation of a system formed after compact binary merger. Our code works with a tabulated equation of state of dense matter, accounts for the neutrino leakage, and follows the mass outflows via the tracer particle method.
We discuss the numerical scheme, and present the recovery method included in our code. We also show results of a numerical simulation, addressed to the post-merger system after the coalescence of binary neutron stars, or a neutron star with a stellar mass black hole. The plasma is very neutron-rich, so the r-process nucleosynthesis in the ejected material may lead to unstable heavy isotopes creation. They are responsible for an electromagnetic signal, observed as a kilonova. In addition, the magnetized, neutrino-driven wind can act as a collimating mechanism for the relativistic jet.}

\FullConference{The European Physical Society Conference on High Energy Physics (EPS-HEP2023)\\
 21-25 August 2023\\
Hamburg, Germany\\}

\begin{document}
\maketitle

\section{Introduction}

Short Gamma Ray Bursts (GRBs) belong to the widely known class of cosmic catastrophic events, which are powered by the relativistic jets ('blast waves') pointing close to the line of sight of the observer. Their spectral energy distributions peak in the gamma-ray band, and the radiation is attributed mainly to the synchrotron process \citep{gehrels_rev}.
The bi-modal duration distribution of GRBs suggests two distinct progenitor types, presumably responsible for their origin. The common feature, which is the relativistic jet, can be powered by the rotational energy of a newly formed black hole, and the process is mediated by the magnetic fields present in the surrounding medium \cite{TNM2011}.
However it is the amount of the mass supply to the engine, and effective build-up of the magnetic field on the black hole horizon, which regulates the engine life-time.

The first proven progenitor  of the short GRB event is known as GW-GRB 170817 and was discovered by the gravitational wave interferometer \citep{abbott2017}. The system of two neutron stars has coalesced and formed a new black hole of about 3 Solar masses, which then launched a jet responsible for the accompanying GRB emission. The time delay of 1.7 s may be associated with formation of a hyper-massive neutron star, prior to the black hole, and the short GRB, powered by accretion of the remnant mass, lasted for about 2 seconds

Another important electromagnetic counterpart to this event was a so-called kilonova \citep{LiPaczynski1998}. The low-energy radiation (in Infrared and Optical bands) was observed for couple weeks after the prompt GRB event \cite{coulter2017}. Characteristic spectral energy distribution proved that the origin of this radiation is a radio-active decay of heavy, unstable isotopes that can be formed in astrophysical dense media via rapid neutron capture (r-process) \cite{kilpatrick2017}.

The dynamical ejecta from compact binary mergers, with mass of about $M_{ej} \sim 0.01 M_{\odot}$, can emit about $10^{40}$-$10^{41}$ erg/s in a timescale of 1 week. Subsequent accretion disk and its outflows, launched after the merger, can provide bluer emission. It can be observed as an additional component in the kilonova lightcurve, and emerge later,  if only it is not absorbed by precedent ejecta \cite{Tanaka2016}.
The 'red' and 'blue' kilonova differ from each other because of different composition of the emitting media \citep{Barnes2016}.


\section{Accretion and outflow simulations}

HARM stands for High Accuracy Relativistic Magneto-hydrodynamics
\cite{gammie2003} and solves the set of hyperbolic equations of GR MHD by the finite volume method, with the constrained transport and classical HLL Riemann solver.
Equation of State (EOS) of an ideal gas with analytic
form was used in the original code, while our group at CTP PAS developed the
code version which uses a relativistic, partially-degenerate Fermi gas EOS, computed numerically and
tabulated. We started from the two-parameter EOS, with $P(\rho, T)$ and $\epsilon(\rho, T)$ implemented in \cite{janiuk2017, janiuk2019}. The neutrino cooling rate was calculated with an approximate two-stream method, accounting for their scattering and absorption via estimation of adequate optical depths.

Below we present the most recent advancement of the code, that self-consistently uses a general three-parameter EOS, and uses the neutrino leakage scheme.


We use the EOS adapted from Helmholtz tables, with $P(\rho, T, Y_{e})$ and $\epsilon(\rho, T, Y_{e})$, where $Y_{e}$ is the electron fraction. It is usable for a wide range of densities and temperatures. The evolving electron fraction is giving an additional source term to the energy equation.

The GR MHD code solves system of equations in the form
\begin{equation}
\partial_{t} U(P) = - \partial_{i} F^{i}(P) + S(P)
\end{equation}
where $U(P)$ is vector of conserved variables, $F^{i}$ are fluxes through cell boundaries, and $S(P)$ are source terms. 
Noticeably, the conserved variables are not the same as those used for the EOS calculation. Instead, in every time-step, the code has to make (typically twice) an inversion between the conserved and primitive variables. There exist a number of inversion schemes, based on specific transformations between these five
independent variables (their explicit form can be found e.g. in \cite{siegel2018}). In this work, we choose the bracketed root-finding method of \cite{palenzuela2015}
which is slower but more robust than other methods, and gives minimum errors for the wide range of temperatures and densities. 


The neutrino leakage scheme computes a gray optical depth estimate along radial rays for electron neutrinos, antineutrinos, and heavy-lepton neutrinos (nux),
and then computes local energy and lepton number loss terms. The scheme is based on equations provided by \cite{rosswog03}.
The source code of the leakage scheme has been downloaded from
{\it https://stellarcollapse.org} and we implemented it in our version of the GR MHD code,
HARM\_COOL\_EOS.

\section{Results}

\begin{figure}
    \includegraphics[width=0.49\textwidth]{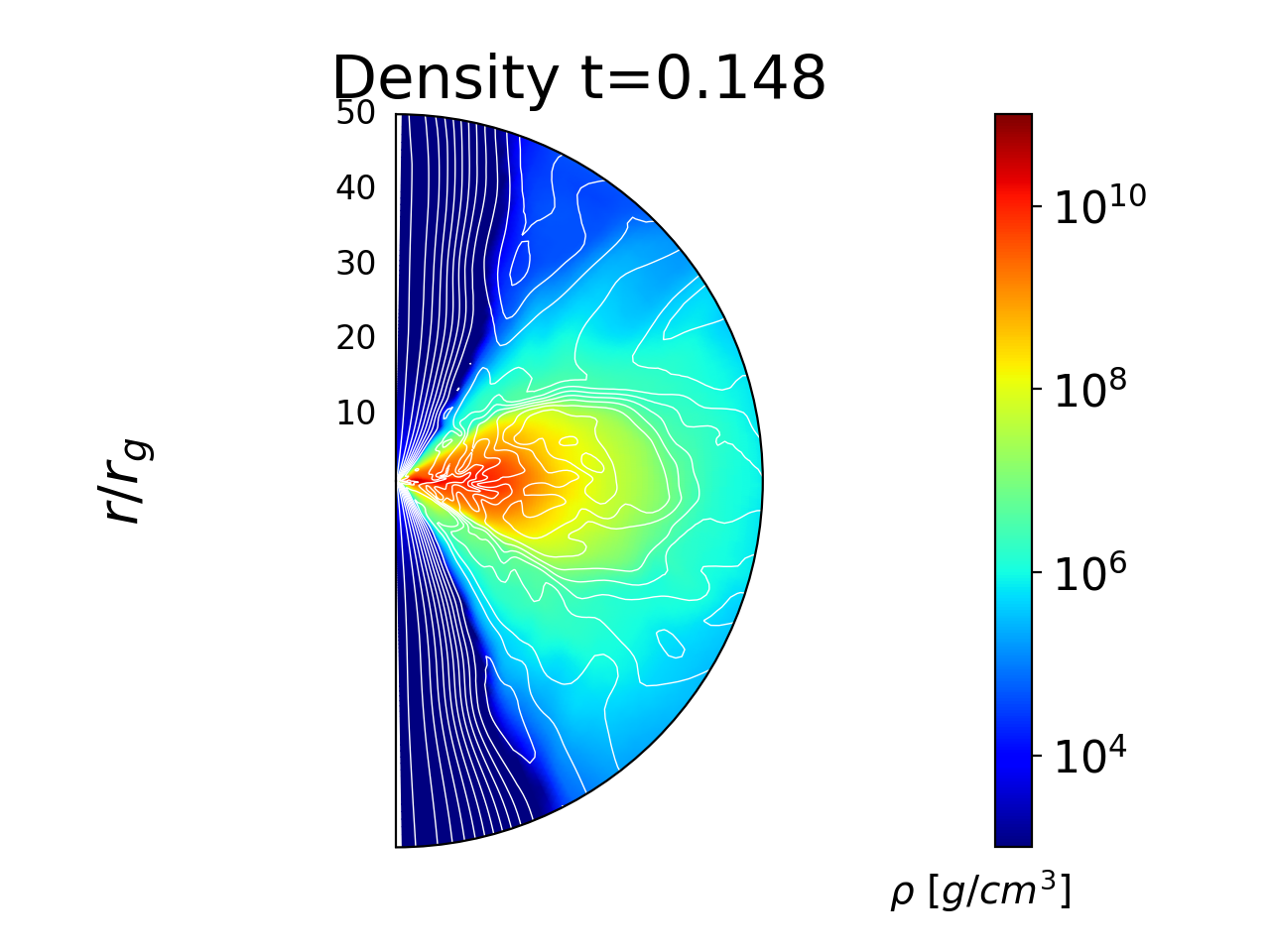}
    \includegraphics[width=0.49\textwidth]{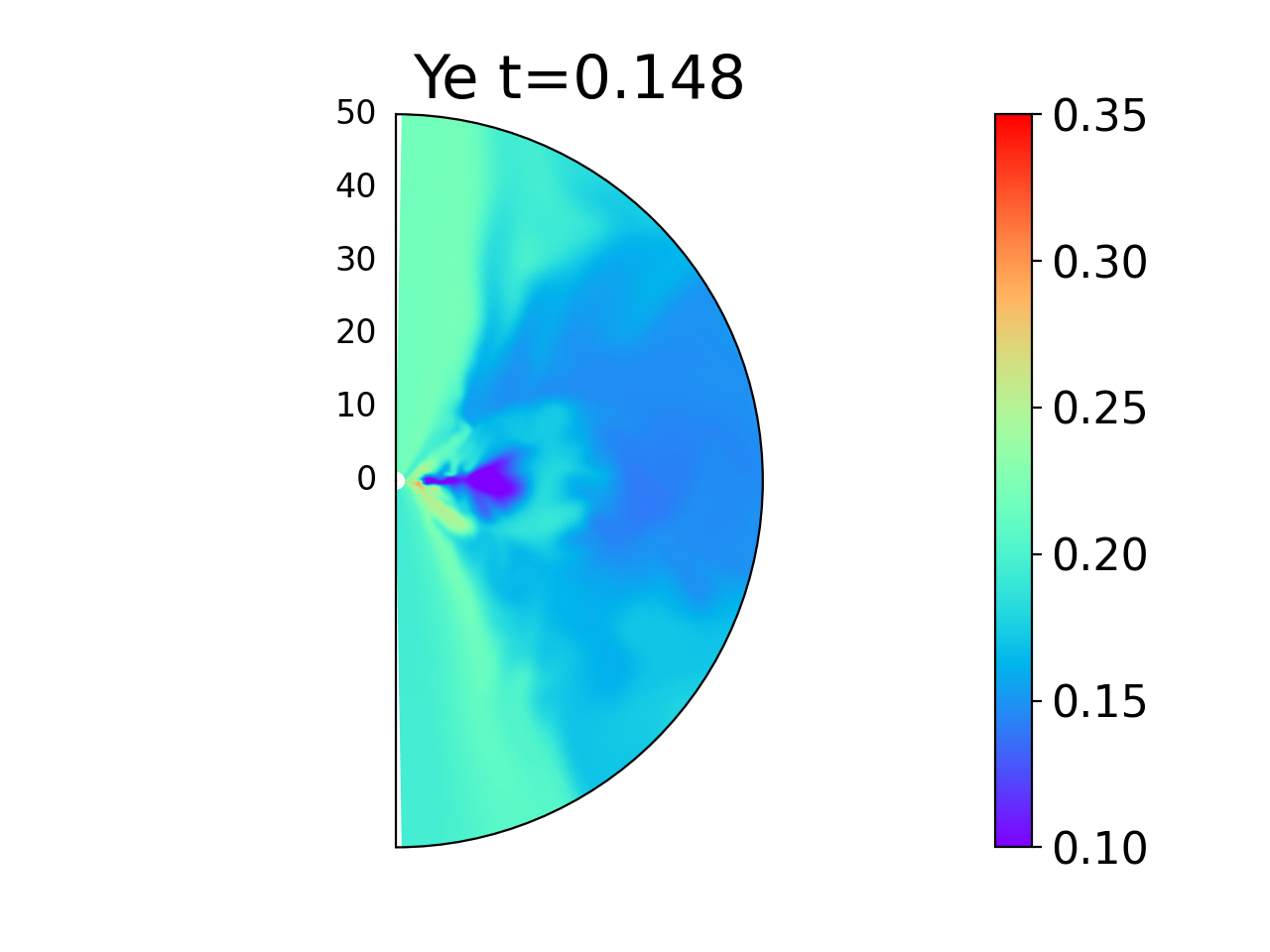}
    \caption{Snapshots from the simulation of the GRB central engine. Left panel shows the density and magnetic field distribution in the innermost part of the accretion disk, at the middle time of the run. The right panel shows its chemical composition parameter - the electron fraction.}
    \label{fig:GRrun}
  \end{figure}

In Figure \ref{fig:GRrun} we show the result of a GR MHD simulation.
Parameters of the model are BH mass of $3 M_{\odot}$, its dimensionless spin $a=0.9375$, and initial gas-to-magnetic pressure ratio $\beta=50$ at the pressure maximum radius, located at $r_{max}=9 r_{g}$.
The disk mass is about $0.1 M_{\odot}$. 
The dense and hot disk launches fast wind outflows ($v/c \sim 0.11-0.23$)
with a broad range of electron
fraction $Y_{e} \sim 0.1-0.4$.
The details of mass loss are sensitive to engine parameters: BH spin and
magnetisation of the disk \cite{nouri2023}.
In general, more magnetized disks produce faster outflows.
These winds contribute to the kilonova
signal, due to radioactive decay of r-process formed isotopes.

\begin{figure}
    \includegraphics[width=0.25\textwidth]{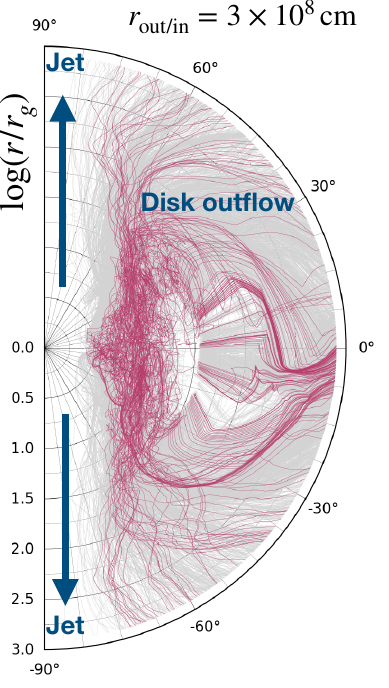}
    \includegraphics[width=0.45\textwidth]{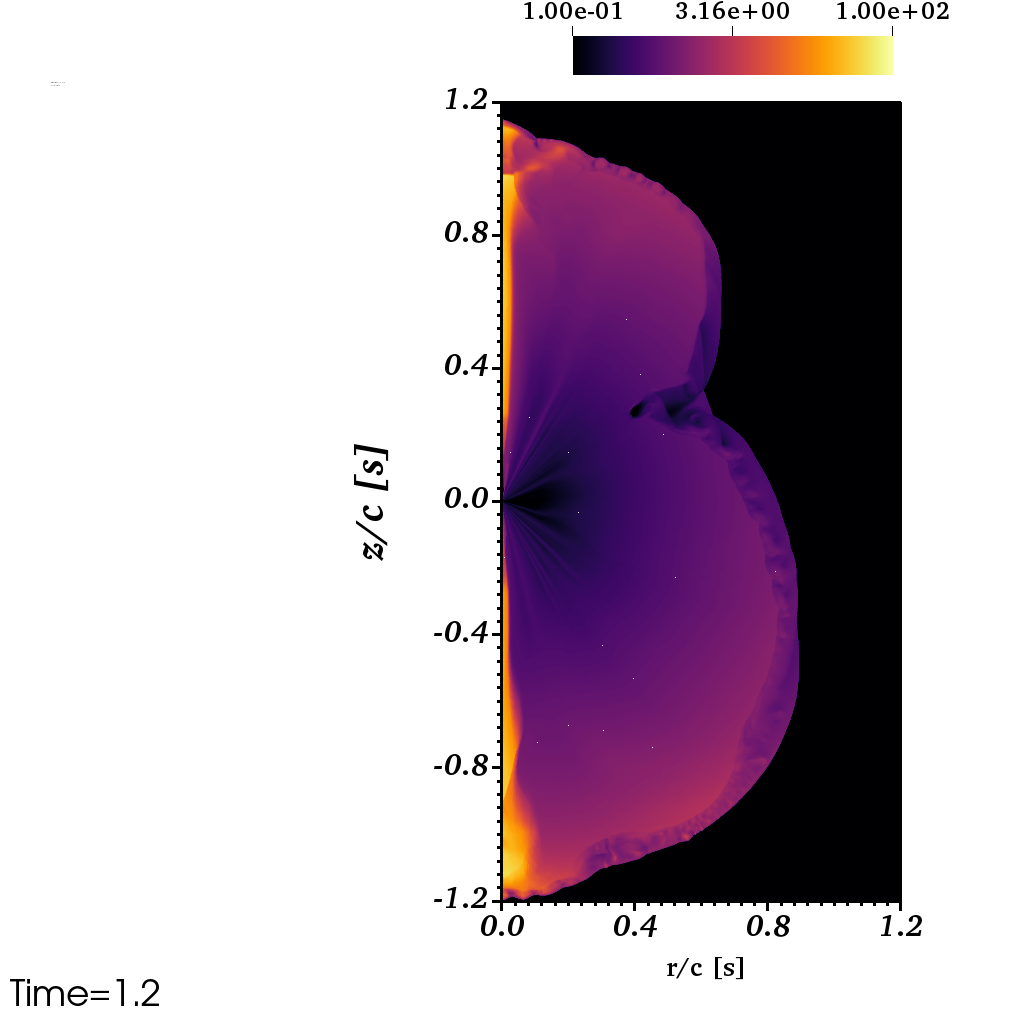}
    \caption{Distribution of tracer particles, following the accretion disk wind launched from the GRB engine (left). The wind inteacts with jet
    launched from the engine after time 1.2 sec, leading to its collimation. Bulk Lorentz factor of the jet is plotted in the color map on the right.}
    \label{fig:tracers}
  \end{figure}

In Figure \ref{fig:tracers}, in its left panel, we show an exemplary run of 2D simulation with tracer particles marking the wind ejecta. Tracers are Lagrangian particles, which store data about density, velocity, and
electron fraction in the outflow, which are important for further r-process nucleosynthesis calculations and estimation of the kilonova lightcurves \cite{nouri2023}. They are initialized uniformly inside the torus, and we typically use about 2000 tracers.
The tracers provide also data for nuclear pressure.
The pressure is taken into account while simulating the jet-wind interaction.
We found that the magnetized,
neutrino-driven wind provides possible collimation mechanism for the GRB jet (Urrutia et al., in prep.).
In the right panel of the plot, we present structure of disk wind and narrow
jet collimated by the wind at its base. This velocity map was
produced with the hydrodynamical simulation which domain is extending three
orders of magnitude further away, outside the jet base region. It was possible thanks to the adaptive mesh refinement technique, as described by \cite{fabio}.

  Our GR MHD code follows the wind outflow, and
the r-process nucleosynthesis is calculated by postprocessing of these data, to obtain chemical evolution of the wind.
The resulting abundance pattern of heavy elements clearly shows the 3 characteristic peaks, corresponding to the mass numbers around $A\sim 60$, $A\sim 130$, and $A\sim 180$. Thus, the process od heavy element formation in the post-merger accretion disk winds is able to explain the chemical composition in the Glactic interstellar medium, including our Solar system, and formation of heaviest elements, such as Gold, Platinium, Uranium, and Thorium.

\section{Conclusions}

Potential electromagnetic counterparts of compact object binary mergers are a
function of the observer viewing angle. Rapid accretion via magnetized, dense and hot disk powers a collimated relativistic jet, which produces the GRB. This event is seen for the on-axis observers only.
Equatorial outflows contribute to lower-energy signal. Both disk wind and jet are powered by the Central Engine: black hole and accretion disk. The wind emission is spherically distributed, therefore it is potentially observable from a wider range of viewing angles. However, as we showed in the simulations, the disk wind acts also as a collimating mechanism for the jet.

The isotopes from the Lanthanide family, known to be formed for $0.15<Y_{e}<0.25$,  are source of higher opacity to the ultraviolet and optical photons, hence the red colour of the emitted radiation. The Lanthanide-free media are producing the other component.
It is therefore essential, that theoretical models of kilonova should not only explain correctly the amount of ejected material and its evolution in time, that govern the kilonova lightcurve peak and decay slope, but also the proper compositions.
We propose that the Lanthanide-free component is associated with the accretion disk wind. In order to verify this hypothesis, we performed simulations of the GRB central engine, and we studied composition changes in the material.

Based on disk wind simulations, we calculated synthetic kilonova ligtcurves for a range of BH-disk mass ratios and range of black hole spin parameters. We find strong correlation between the BH spin and ejected mass. Our models do not provide direct method to distinguish between BH-NS and NS-NS
progenitors, but the slopes of kilonova signal, as well as jet properties, may sligtly differ between these two types of progenitor.

\begin{figure}
    \includegraphics[width=0.5\textwidth]{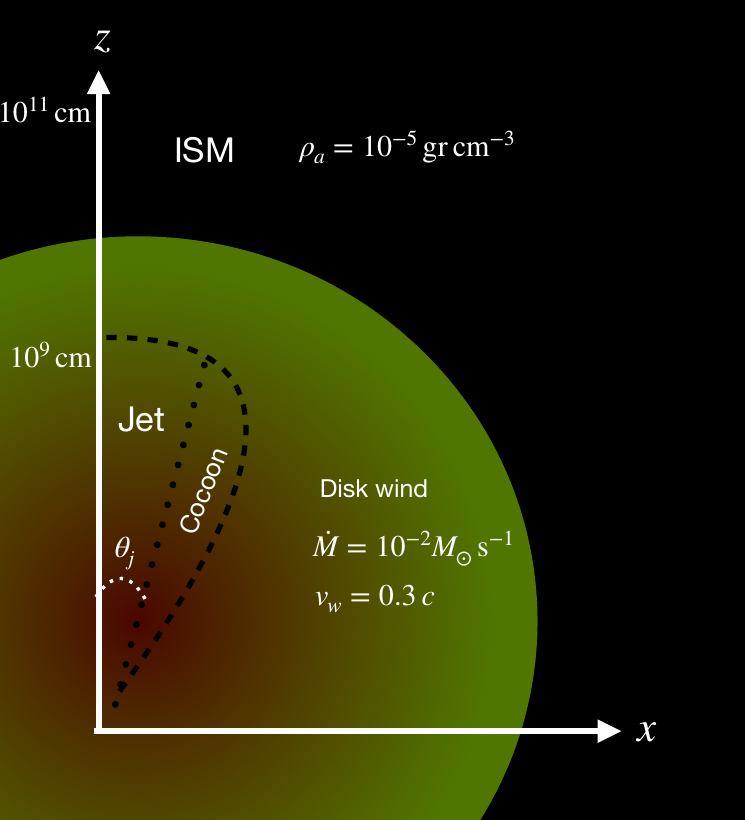}
    \caption{Schematic view of the disk wind interacting with relativistic jet, in a post-merger system. The ambient medium density is much smaller than that of the wind.}
    \label{fig:cartoon}
  \end{figure}

In Figure \ref{fig:cartoon} we show a schematic view of the system formed after the compact binary merger.
The plot is highlighting the range of physical scales involved.
The accretion disk structure, neutrino emission, mass loss rate and mildly relativistic velocities of wind ejecta are determined from our GR MHD model. The jet, assumed to be launched from the same central engine, and reaching much larger distances than diluted disk wind, is further propagating in a flat space, filled with the ambient low density (ISM) medium, uniformly distributed.
The density and pressure of the wind at the base pushes a weak 'cocoon', and in consequence it collimates the jet through the pressure balance.
We conclude that complex system formed after the compact binary merger and the interaction of relativistic jet with the wind ejecta shape the structure of outflow and its radiation properties.


\begin{thebibliography}{99}
   
\bibitem{abbott2017} B.P. Abbott et al., {\it Gravitational Waves and Gamma-Rays from a Binary Neutron Star Merger: GW170817 and GRB 170817A}, \emph{ApJL} {\bf 848} 13 (2017)
\bibitem{Barnes2016} J. Barnes, {\it Radioactivity and Thermalization in the Ejecta of Compact Object Mergers and Their Impact on Kilonova Light Curves}, \emph{ApJ} {\bf 829} 110 (2016)
\bibitem{coulter2017} D.A. Coulter, et al.,   {\it Swope Supernova Survey 2017a (SSS17a), the optical counterpart to a gravitational wave source  }, \emph{Sci} {\bf 358} 1556 (2017)
  \bibitem{fabio} F. De Colle, et al., {\it Gamma-Ray Burst Dynamics and Afterglow Radiation from Adaptive Mesh Refinement, Special Relativistic Hydrodynamic Simulations}, \emph{ApJ} {\bf 746} 122 (2012)
\bibitem{gammie2003} C.F. Gammie, J.C. McKinney, G. Toth, {\it HARM: A Numerical Scheme for General Relativistic Magnetohydrodynamics}, \emph{ApJ} {\bf 589} 444 (2003)
\bibitem{gehrels_rev} N. Gehrels, E. Ramirez-Ruiz, D.B. Fox, {\it Gamma-Ray Bursts in the Swift Era}, \emph{Ann. Rev. Astron. \& Astroph.} {\bf 47} 567 (2009)
\bibitem{janiuk2017} A. Janiuk, {\it Microphysics in the Gamma-Ray Burst Central Engine}, \emph{ApJ} {\bf 837} 39 (2017)
  \bibitem{janiuk2019} A. Janiuk, {\it The r-process Nucleosynthesis in the Outflows from Short GRB Accretion Disks}, \emph{ApJ} {\bf 882} 163 (2019)
\bibitem{kilpatrick2017} C.D. Kilpatrick, et al., {\it Electromagnetic evidence that SSS17a is the result of a binary neutron star merger}, \emph{Sci} {\bf 358} 1583 (2017)
 \bibitem{LiPaczynski1998} L.X. Li, Paczynski B., {\it Transient Events from Neutron Star Mergers}, \emph{ApJL} {\bf 507} 59 (1998)
 \bibitem{nouri2023} F.H. Nouri, A. Janiuk, M. Przerwa, {\it Studying Postmerger Outflows from Magnetized-neutrino-cooled Accretion Disks}, \emph{ApJ} {\bf 944} 220 (2023)
   \bibitem{palenzuela2015} C. Palenzuela, et al.,  {\it Effects of the microphysical equation of state in the mergers of magnetized neutron stars with neutrino cooling} \emph{Phys Rev. D.} {\bf 92} 044045 (2015)
 \bibitem{rosswog03} S. Rosswog, M. Liebendoerfer, {\it High-resolution calculations of merging neutron stars - II. Neutrino emission}, \emph{MNRAS} {\bf 342} 673 (2003)
   \bibitem{siegel2018} D. Siegel, et al., {\it Recovery Schemes for Primitive Variables in General-relativistic Magnetohydrodynamics} \emph{ApJ} {\bf 859} 71 (2018)
\bibitem{Tanaka2016}  M. Tanaka, {\it Kilonova/Macronova Emission from Compact Binary Mergers}, \emph{Adv. Astron.}, {\bf 634197} (2016)
\bibitem{TNM2011}  A. Tchekhovskoy, R. Narayan, J.C. McKinney,  {\it Efficient generation of jets from magnetically arrested accretion on a rapidly spinning black hole} \emph{MNRAS Lett.} {\bf 418} 79 (2011)

\end{thebibliography}
\end{document}